\documentclass{aastex}       

\usepackage{graphicx,times}             
\usepackage{subfigure}
\usepackage{natbib}
\usepackage{textcomp}
\usepackage{mathtext}
\usepackage{float}
\usepackage{amssymb,amsmath}
\usepackage{longtable,booktabs}



\newcommand{\fermi}{\textit{Fermi}}
\newcommand{\gr}{$\gamma$-ray}

\begin{document}


\slugcomment{Accepted for publication of Research in Astronomy and Astrophysics}

\title{Searching for $\gamma$-ray Counterparts to Very Faint X-Ray Transient Neutron Star binaries}

\author{Gege Wang $^1$ $^2$ \and Zhongxiang Wang$^1$} 
\affil{$^1$Shanghai Astronomical Observatory, Chinese Academy of Sciences,
             Shanghai 200030, China; {\it wangzx@shao.ac.cn}}
\affil{$^2$Graduate University of the Chinese Academy of Sciences, No. 19A, Yuquan Road, Beijing 100049, China}

\begin{abstract}
Very faint X-ray transients (VFXTs) are a group of X-ray binaries with 
low luminosities, the peak X-ray luminosities during their outbursts being only 
10$^{34}$--10$^{36}$ erg\,s$^{-1}$. Using the $\gamma$-ray data obtained with
the Large Area Telescope (LAT) onboard {\it Fermi Gamma-Ray Space Telescope}, 
we investigate their possible nature of containing rotation-powered pulsars, 
or more specifically being transitional millisecond pulsars (MSPs).
Among more than 40 known VFXTs, we select 12 neutron star systems. We analyze
the LAT data for the fields of the 12 VFXTs in 0.2--300 GeV energy range, but
do not find any counterparts likely detected by {\it Fermi}. We obtain
the luminosity upper limits for the 12 sources. While the distances to
the sources are largely uncertain, the upper limits are comparable to
the luminosities of two transitional systems PSR J1023$-$0038 and 
XSS J12270$-$4859. From our study, we conclude that no evidence is found
at $\gamma$-rays for the suggestion that some of VFXTs could contain
rotation-powered MSPs (or be transitional MSP systems).
\end{abstract}




\section{Introduction}
\label{sect:intro}

The discovery of the millisecond pulsar (MSP) in the low-mass X-ray 
binary (LMXB) SAX J1808.4$-$3658 \citep{wv98} has provided long-sought 
evidence for the evolution scenario that MSPs form from neutron star
LMXBs \citep{alp+82,rs82}. The multi-wavelength properties of this so-called
accretion-powered MSP system actually suggest that in its quiescent state
when its X-ray luminosity is $\sim 5\times 10^{31}$ erg\,s$^{-1}$, the
neutron star switches to be a rotation-powered pulsar (\citealt{bur+03};
see also \citealt{wan+13} and references therein). The similar feature has
also been suggested to other accretion-powered MSP systems (e.g.,
\citealt{dav+07,dav+09}). Then discovery of the transitional MSP 
binary PSR J1023+0038 \citep{arc+09} has helped make the picture clearer. This
system has shown that it can switch between the states of
being a normal radio pulsar binary and having an accretion disk around
the neutron star (e.g., \citealt{arc+09,wan+09,sta+14,tak+14}). More recently,
the neutron star binaries J1824$-$2452I \citep{pap+13} and XSS J12270$-$4859
\citep{bas+14,dem+14}, where the first is located in the globular cluster M28,
have also been found to be able to switch between the states. The feature
of the transitional pulsar binaries shows again that at the end of LMXB
evolution, a system can have both a rotation-powered MSP and an accretion
disk.

Based on observational studies of these systems, it can be seen that
they share the similar property of having low X-ray luminosities. For example,
even in the active state with an accretion disk, PSR J1023+0038 has an
X-ray luminosity of only $\sim 6\times 10^{33}$ erg s$^{-1}$ \citep{li+14}. 
The low luminosity implies a low accretion rate in the disk,
and thus the energy output from the neutron star powered by its fast rotation
can distroy the weak accretion disk (see, e.g., \citealt{wan+09}). X-ray
observations of PSR J1023+0038 in its disk-free state has clearly shown 
the existence of an intrabinary shock region that is the result of 
the interaction between the pulsar wind and the outflow of the companion star
\citep{bog+11}. The mass transfer rate must also be low such that 
the outflow from the companion can be stopped from forming an accretion
disk by the pulsar wind for years. Following this idea, we may consider
that LMXB sources, identified because of the detection of them from 
X-ray surveys, may contain rotation-powered pulsars. For example, both 
PSR J1023+0038 and XSS J12270$-$4859 in their active state were once 
considered to be typical LMXBs \citep{ta05,hil+11}.

Because of low peak X-ray luminosities (10$^{34}$--10$^{36}$ erg\,s$^{-1}$)
during their outbursts, a group of X-ray binaries are classified as very faint
X-ray transients (VFXTs; \citealt{wij+06}). While their nature of being
so faint are poorly understood (e.g., \citealt{hei+15}), observations in
the past 10 years have established that a fraction of the VFXTs are neutron 
star LMXBs, due to the detection of thermonuclear type I X-ray bursts 
(see, e.g., \citealt{cam09}). It has thus been pointed out that some of 
these neutron star binary
systems could correspond to the active state of the transitional MSP systems
\citep{hei+15}. Given that both PSR J1023+0038 and XSS J12270$-$4859 have
shown significant $\gamma$-ray emission during their active states (actually
their $\gamma$-ray fluxes were several times brighter than those in their
disk-free state), $\gamma$-ray observations of the VFXTs can be
used to explore this possibility.

The \textit{Fermi Gamma-Ray Space Telescope (Fermi)}, which was launched 
in 2008 June, has a Large Area Telescope (LAT) onboard. The unprecedented
sensitivity of the LAT at $\gamma$-rays has allowed us to study both
Galactic and extra-galactic high-energy sources in detail.
Thus far, more than 3000 sources have been detected by the LAT \citep{3fgl15}. 
In this paper, we report our search for the $\gamma$-ray
counterparts to 12 neutron star VFXTs (see Table~1) from the analysis of 
the LAT all-sky 
survey data. These VFXTs were selected from more than 40 identified VFXTs, 
a list of which are summarized in \citet{ll17}. In our selection, we avoided
the Galactic center and globular cluster sources; the extremely high source
density in the regions do not allow a clear identification of any faint
sources in the {\it Fermi} data. 

\bigskip

\section{\fermi\ LAT Data Analysis and Results}
\label{sec:ar}
\subsection{\fermi\ LAT Data}
 
As one of the two instruments onboard \textit{Fermi}, LAT is an 
imaging \gr\ telescope scanning over the whole sky 
every three hours in the energy range from 20 MeV to 300 GeV
\citep{atw+09}.  In the analysis reported in this paper, 
the LAT data for each target were selected from the latest \textit{Fermi} 
Pass 8 database within 20 $^\circ$ of the target's position. 
The observing time period of the data was from 2008 August 4 15:43:39 
to 2016 November 1 00:00:00 (UTC), nearly 8.25 years. 
To avoid the relative large uncertainties of the instrument response 
function of the LAT in the low energy range,
the energy range we used was from 200 MeV to 300 GeV.  
In addition, following the recommendations of 
the LAT team\footnote{http://fermi.gsfc.nasa.gov/ssc/data/analysis/scitools/}, 
we selected events with zenith angles less than 90 deg to exclude the possible
Earth's limb contamination.

\subsection{Maximum Likelihood Analysis}

Using the newly released LAT science tools package {\tt v10r0p5},
we conducted standard binned maximum likelihood analysis \citep{mat+96} 
to the data for each target.
The source model of a target contained all the sources,
based on the LAT 4-year catalog \citep{3fgl15}, within 20 deg 
centered at the target's position. 
The normalization parameters and the spectral indices of the 
sources within 5 deg from each target were set as free parameters
and all the other parameters were fixed at their catalog values in
the LAT catalog. We included
the Galactic and the extragalactic diffuse emission by using
the model gll\_iem\_v06.fits and spectrum file iso\_P8R2\_SOURCE\_V6\_v06.txt 
in the source model, respectively. The normalization parameters of 
the two diffuse emission components were set as free paramters. 
For each of the 12 VFXT targets, we set a simple power law in the source model. 

We calculated Test Statistic (TS) maps of a $3^\circ\times 3^\circ$ region 
centered at the position of each target. TS values are derived from 
TS$=-2\log (L_{0}/L_{1})$, where $ L_{0} $ and $ L_{1} $ are the maximum 
likelihood values for a model having none or an additional source 
at a given position respectively \citep{abd+10}. 
The TS value for a detected source is approximately equal to the square of 
the detection significance \citep{abd+10}. 

\begin{figure}[H]
   \centering
   \includegraphics[width=0.42\textwidth, angle=0]{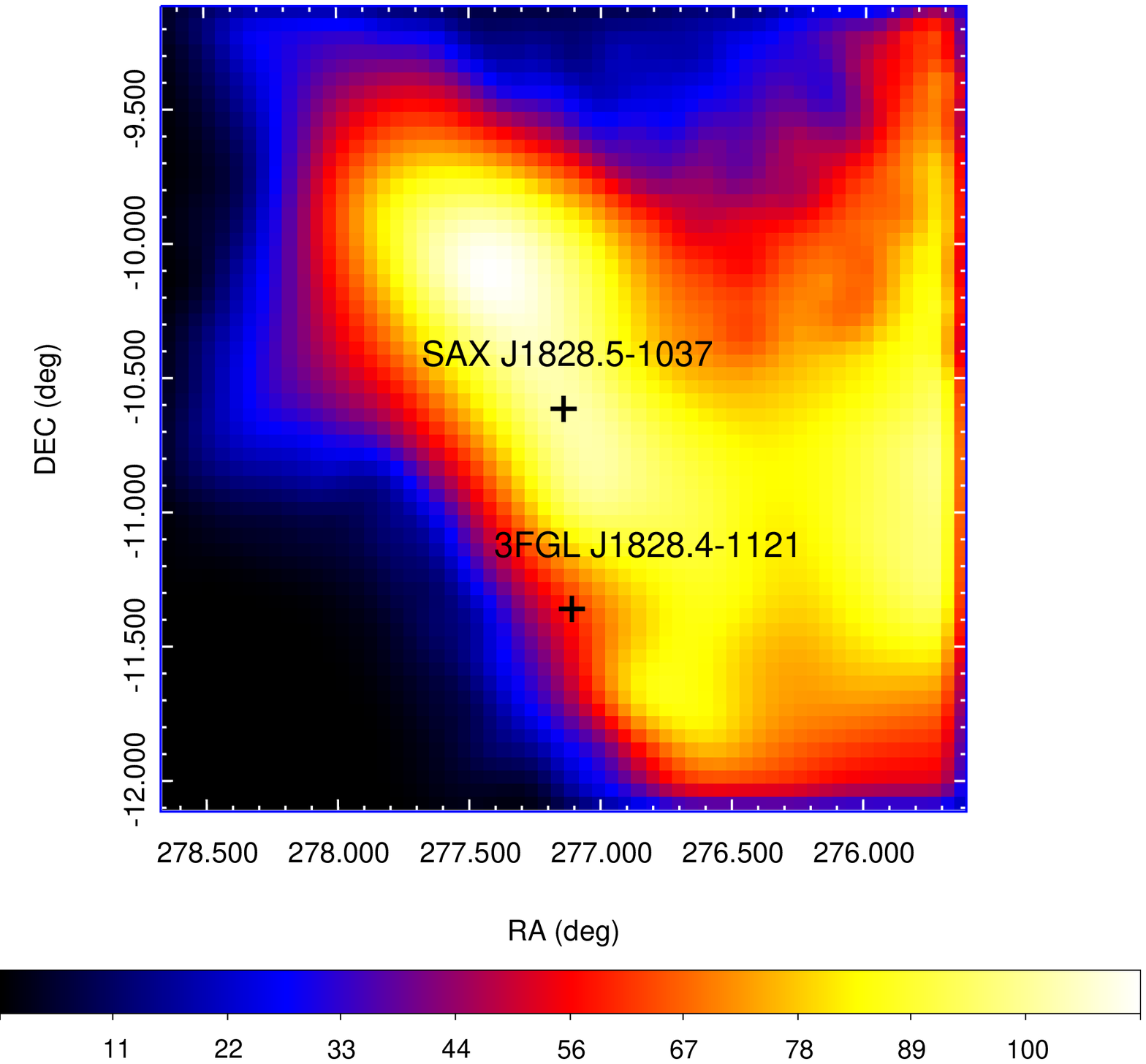}
   \includegraphics[width=0.45\textwidth, angle=0]{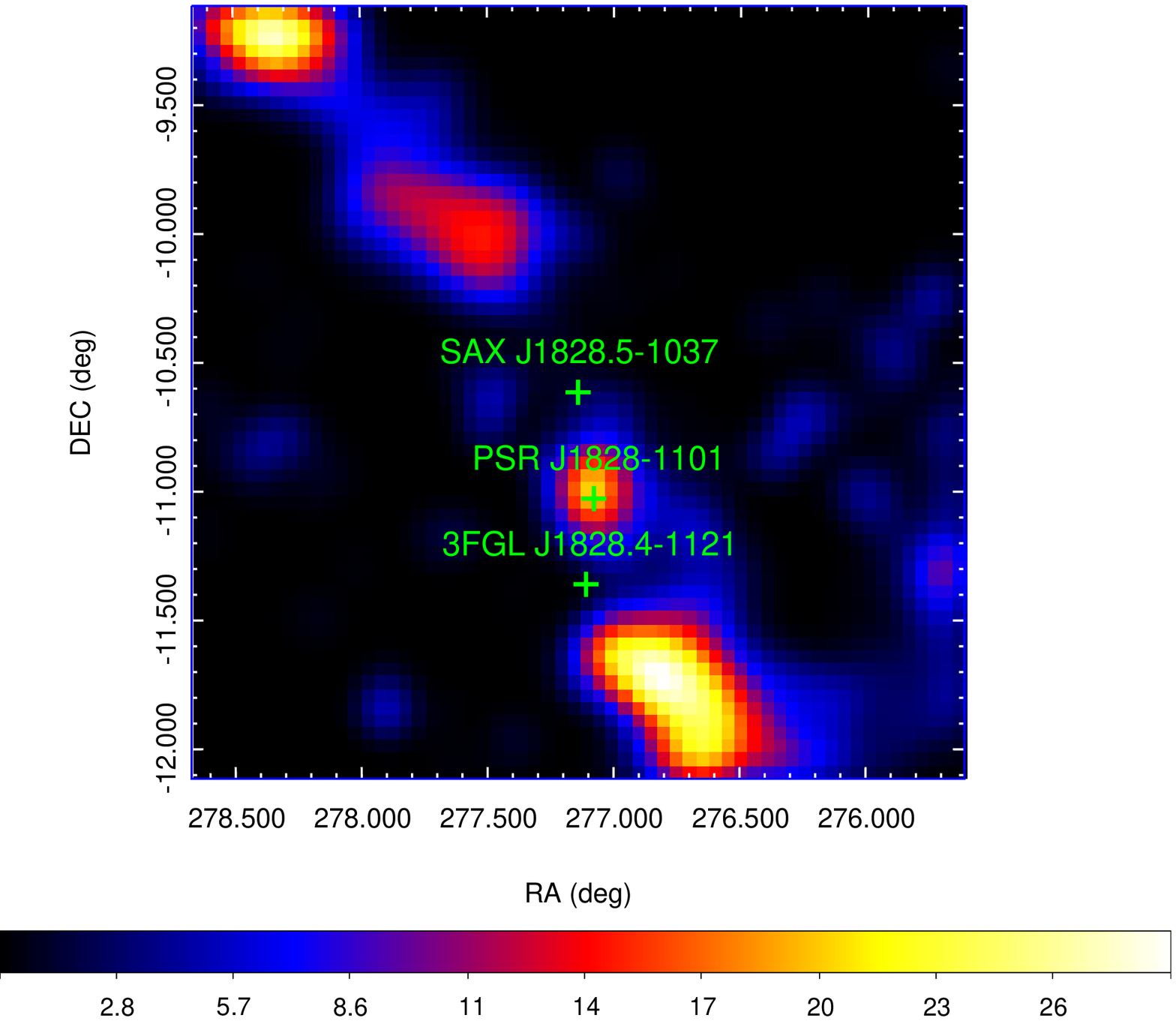}
   \caption{TS maps for SAX J1828.5$-$1037 in the energy range of
0.2--300 GeV ({\it left panel}) and 1--300 GeV ({\it right panel}).
The image scale is 0.05${^\circ}$ pixel$^{-1}$. 
The positions of the LAT third source catalog source (3FGL) J1828.4$-$1121,
PSR J1828$-$1101, and SAX J1828.5$-$1037 are marked with crosses.}
   \label{fig:ts}
\end{figure}

In the TS maps we obtained, an extended region with high TS values was
detected around the target SAX~J1828.5$-$1037 (see the left panel of 
Figure~\ref{fig:ts}). For all other targets, no such regions or possible
point-like sources with sufficiently high TS values were seen. We further
analyzed the data for SAX~J1828.5$-$1037 by calculating a 1--300 GeV TS map.
In this higher energy range, the spatial resolution of LAT is significantly 
improved
to be approximately 1 deg (68\% containment angle; see details in the LAT performance 
note\footnote{http://www.slac.stanford.edu/exp/glast/groups/canda/lat\_Performance.htm}).
The TS map is shown in the right panel of Figure~\ref{fig:ts}, and it can be 
seen that there is no $\gamma$-ray source at the source's position and
the high TS values at low energy ranges are likely caused by the pulsar
J1828$-$1101 south to SAX~J1828.5$-$1037.

\subsection{Variability Search}

All these targets are variable sources in the X-ray band. For example, X-ray outbursts were
reported in SAX J1753.5$-$2349 \citep{del+10}, 
SAX J1806.5$-$2215 \citep{alt+11}, and
Swift J185003.2$-$005627 \citep{deg+12} after {\it Fermi} was launched. 
We therefore searched for 
detections considering if they had significant variations during 
the {\it Fermi} LAT coverage. For each target, we divided the whole 8.25 years
data into bins with two time durations, 30 days and 180 days, 
and performed likelihood analysis to the data in each time bin. However
the TS values at the position of each target did not show any possible 
detections. We concluded that no target was detected
in the time bins of either 30 days or 180 days.

\subsection{Luminosity Upper Limit Estimation}

We estimated the \gr\ flux upper limits of the 12 VFXTs, performing the 
binned likelihood analysis with the source models described above. 
The spectral normalization factors were set as free parameters. 
Following the procedure given by the LAT team, 
we obtained the 95\% flux upper limits in 0.2--300 GeV energy range
by increasing the flux values until 
the maximum likelihood values decreased by $e/2$ in logarithm. 
In the calculations, the photon index $\Gamma$ of the power law was set 
the same as that of the transitional MSP binary PSR J1023+0038 in its 
disk-free state, which is 2.4. 
While the distances to the VFXTs are highly uncertain, most of them have
upper limits, estimated by setting the burst peak luminosities equal
to the Eddington limit (e.g., \citealt{cam09}).    
In Table~\ref{tab:tabone}, the distance values summarized by \citet{ll17}
are given. Using the values, we obtained the upper limits on
the \gr\ luminosities, which are provided in Table~\ref{tab:tabone}.
Note that because of different assumptions for the bursts, three sources 
have two different distance upper limits. We always used the larger values.

\section{Discussion}
\label{sec:rd}

Having selected 12 neutron star systems from more than 40 known VFXTs,
we have investigated their possible nature at \gr\ energies. The data were
from {\it Fermi} LAT all-sky survey at the energy range of 0.2--300 GeV.
The motivation was that if some of neutron star VFXTs are the type of
X-ray binaries similar to accretion-powered MSP or transitional MSP
binaries, we would expect possible detection of them at $\gamma$-rays since
they could switch to be rotation powered in quiescence
or have bright \gr\ emission such as in the active state of 
the transitional MSP binaries.
However from our analysis of 8.25-year LAT data, we did not detect
any possible candidates. We note that nearly all of these sources, 
except SAX J1818.7+1424, are located in the Galactic disk 
(Galactic latitudes smaller than 5 deg) and most of them are close to 
the direction of the Galactic center (due to the designed X-ray monitoring; 
e.g., \citealt{wij+06}).
Our search was hampered by the crowdedness in the fields. There are
often several bright sources in a source field of $3^\circ\times 3^\circ$ we
considered (such as in Figure~\ref{fig:ts}), and they made a clear search 
for any faint sources difficult.
\begin{figure}[H]
   \centering
   \includegraphics[width=0.72\textwidth, angle=0]{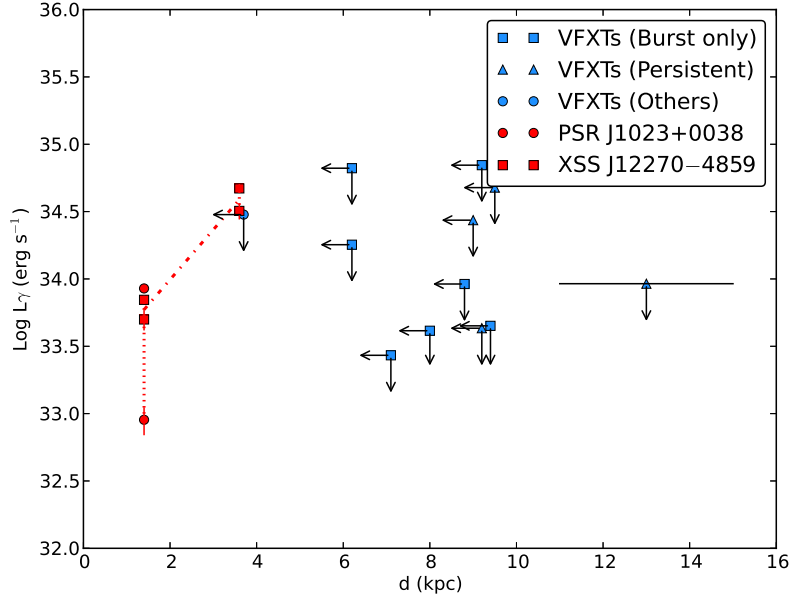}
   \caption{Luminosity upper limits in 0.2--300 GeV for the 12 VFXTs. Also shown
are the luminosity ranges of the transitional MSP binaries PSR J1023+0038 
(red circles) and XSS J12270$-$4859 (red squares) between the active and
disk-free states. Because XSS J12270$-$4859 is known with a distance
range of 1.4--3.6 kpc, we show both luminosity ranges at 1.4 and 3.6 kpc.}
   \label{fig:two}
\end{figure}

Using the estimated upper limits on the \gr\ luminosities of the VFXTs
in 0.2--300 GeV,
we can compare them with the two transitional MSP binaries 
PSR J1023+0038 and XSS J12270$-$4859, whose $\gamma$-ray emission was
relatively well studied by {\it Fermi} LAT.
In Figure~\ref{fig:two}, we show the upper limits for the VFXTs and
the luminosity ranges of the two transitional MSP binaries. 
Note that for most VFXTs, their distances only have upper limits.
Since in the active state of the latter two sources, their \gr\ luminosities 
were significantly brighter \citep{sta+14,xw15}, we plotted both luminosities
in the disk-free and active states.
In addition, the distance to XSS J12270$-$4859 is rather uncertain 
(e.g., \citealt{roy+15,riv+17}), and we considered the range of 1.4--3.6 kpc. 
The comparison shows that the upper limits of the VFXTs are actually
comparable to the luminosity ranges of the two transitional MSPs. 
For XSS J12270$-$4859,
in the disk-free state even assuming the low distance value of 1.4~kpc,
a few of the VFXTs' upper limits are already lower than its luminosity.
For
PSR J1023$+$0038, its disk-free luminosity was $\simeq$10$^{33}$ erg\,s$^{-1}$
\citep{tam+10}, and the upper limits have not reached such a low level.

Among the 12 VFXTs, seven are classified as burst-only 
sources (Table~\ref{tab:tabone}; see also \citealt{ll17}), which
implies that they had extremely low persistent X-ray emission (X-ray luminosites
lower than 10$^{33}$ erg\,s$^{-1}$). There are four 
persistent sources, which are defined because they have X-ray luminosities 
at the level of $\sim$10$^{34}$ erg\,s$^{-1}$. In order to explain
temporal properties of VFXTs' X-ray emission, \citet{hei+15} have suggested 
that persistent VFXTs could be transitional MSPs in the active state. 
In Figure~\ref{fig:two}, we can see that one of the persistent sources has
the upper limit lower than the active-state luminosities of the
both transitional MSPs.  In addition, three burst-only sources have
the upper limits below the luminosity of XSS J12270$-$4859 in its disk-free
state (during which radio pulsed emission has been detected; \citealt{roy+15}).
From the comparison, we may conclude that no evidence is found at $\gamma$-rays
to support VFXTs contain pulsars. However we should be 
cautious that pulsars have a large range of $\gamma$-ray luminosity values due
to different properties of the pulsars or geometric effects \citep{2fpsr13}, 
and even for the active state of the transitional MSPs, how their enhanced
$\gamma$-ray emission is produced is still not clear \citep{pt15}.

\begin{acknowledgements}
This research made use of the High Performance Computing Resource in the Core
Facility for Advanced Research Computing at Shanghai Astronomical Observatory.
This research was supported by the National Program on Key Research 
and Development Project (Grant No. 2016YFA0400804) and
the National Natural Science Foundation of China (11633007). 
Z.W. acknowledges the support by the CAS/SAFEA International Partnership 
Program for Creative Research Teams.
\end{acknowledgements}

\begin{table}
\begin{minipage}[]{100mm}
 \caption{Luminosity upper limits in 0.2--300 GeV for the 12 VFXTs. }
 \label{tab:tabone}
\end{minipage}
\tiny
\centering
\begin{tabular*}{0.95\textwidth}{@{\extracolsep{\fill} } lccccc}
\toprule\toprule
Source Name & R.A. & Dec. & Distance & $L_{\gamma}/10^{35}$ & Type \\
\qquad &  &  & (kpc) & (erg s$^{-1}$) & \\
\midrule
SAX J1324.5$-$6313  &  13:24:27.00  &  -63:13:24.00  &  $<$ 6.2  &  0.2 & Burst-only  \\
SAX J1753.5$-$2349  &  17:53:31.90  &  -23:49:14.86  &  $<$ 8.8  &  0.09   & Burst-only\\
SAX J1752.3$-$3138  &  17:52:24.00  &  -31:37:42.00  &  $<$ 9.2  &  0.7   & Burst-only\\
SAX J1806.5$-$2215  &  18:06:32.18  &  -22:14:17.20  &  $<$ 8.0  &  0.04   & Burst-only\\
SAX J1818.7+1424  & 18:18:44.00  &  +14:24:12.00  &  $<$ 9.4  & 0.04   & Burst-only \\
SAX J1828.5$-$1037  &  18:28:33.00  &  -10:37:48.00  &  $<$ 6.2  &  0.7   & Burst-only\\
SAX J2224.9+5421  &  22:24:52.00  &  +54:21:54.00  &  $<$ 7.1  &  0.03    & Burst-only\\
1RXS J170854.4$-$321857  &  17:08:52.50  &  -32:19:26.00  &  13$\pm$2.0 &  0.09  & Persistent \\
RX J1718.4$-$4029  &  17:18:24.13  &  -40:29:30.40  &  $<$ 6.5 or  $<$ 9.0 &  0.3  & Persistent\\
RX J1735.3$-$3540  &  17:35:23.75  &  -35:40:16.00  &  $<$ 6.2 or  $<$ 9.5  &  0.5  & Persistent\\
AX J1754.2$-$2754  &  17:54:14.47  &  -27:54:36.10  &  $<$ 6.6 or  $<$ 9.2 & 0.04  & Persistent\\
Swift J185003.2$-$005627  &  18:50:03.33  &  -00:56:23.30  &  $<$ 3.7  &  0.3   & Others \\
\bottomrule
\end{tabular*}
\label{tab:tabfirst}
\end{table}

\end{document}